\documentclass[aps,twocolumn]{revtex4}
\usepackage{graphics}
\usepackage{graphicx}
\usepackage{amsmath}
\usepackage{amssymb}
\usepackage{float}

\providecommand{\ket}[1]{|#1\rangle}
\providecommand{\bra}[1]{\langle#1|}
\providecommand{\kebra}[2]{\ket{#1}\bra{#2}}

\begin{document}
\title{Qubit Mediated Time Robust Entangling of Oscillators in Thermal Environments}
\author{Tommaso Tufarelli$^1$,  M. S. Kim$^2$, Sougato Bose$^1$}
\affiliation{$^1$Department of Physics and Astronomy, University College London, Gower Street, London WC1E 6BT, UK\\
$^2$School of Mathematics and Physics, Queen's University Belfast, University Road, Belfast BT7 1NN, UK}
\begin{abstract}
We consider two separated oscillators initially in equilibrium and continuously interacting with thermal environments, and propose a way to
entangle them using a mediating qubit. An appropriate interaction allows for an analytic treatment of the open system, removes the necessity of
fine-tuning interaction times and results in a high tolerance of the entanglement to finite temperature. The entanglement thus produced between
the oscillators can be verified either through a Bell inequality relying on oscillator parity measurements or through conditional extraction of
the entanglement on to two mutually non-interacting qubits. The latter process also shows that the generated mixed entangled state of the
oscillators is an useful resource for entangling qubits. By allowing for influences from environments, taking feasible qubit-oscillator
interactions and  measurement settings, this scheme should be implementable in a variety of experimental setups. The method presented for the solution of the master equation can also be adapted to a variety of problems involving the same form of qubit-oscillator interaction
\end{abstract}

\maketitle
\section*{INTRODUCTION}
Two of the simplest quantum systems are the qubit and the harmonic oscillator (oscillator for brevity). Their controlled interaction is realizable in a multitude of
settings. These include cavity-quantum electrodynamics (QED), where atomic qubits interact with electromagnetic fields \cite{Haroche1},
circuit-QED, where superconducting qubits interact with microwave resonators \cite{Blais} or quantum nano-mechanics where mechanical
oscillators interact with superconducting qubits \cite{LaHaye} and atomic qubits \cite{Treutlein}. It is generally perceived that the oscillators will have to be first cooled
to a known pure state and isolated sufficiently from their environments before quantum effects in them can be seen. However, one can ask the following question in order to lower the demands for controlling quantum systems. How far can cooling and isolation be relaxed? We answer this
question in the context of entangling two oscillators by a mediating qubit. This is important for those qubits which do not interact with each
other. The entanglement of the oscillators could then be a resource for entangling qubits and their teleportation
\cite{Davidovich,Browne,Bose-Kim,Palma-Kim}. Additionally, entangling mesoscopic oscillators, as available in quantum nanomechanics, fundamentally probes the
boundary of the quantum world.

  The most natural initial state of an oscillator is a {\em thermal
state}. Moreover an oscillator {\em continuously interacts with
its
  thermal environment} even while interacting with a qubit. Can
one use the available qubit-oscillator interactions to entangle two oscillators in the above situation in a verifiable way? Surprisingly, though work on
entangling oscillators initially in ``pure states" is aplenty \cite{Palma-Kim,Davidovich,Browne,Agarwal-Bose,Agarwal-Solano}, only very recently the
first major step in answering the above question has been taken. Refs.\cite{Ralph,Zheng} have shown how to entangle two thermal state oscillators by a mediator in a
pure state with some heuristic/approximate considerations of decoherence induced by zero temperature baths (as opposed to a thermal environment). However,
wherever thermal states are relevant, the thermal environment is also {\em unavoidable}. An exact analytic treatment of this problem by solving the appropriate master equation
for a thermal environment is required to go beyond heuristics and ascertaining how the openness of the oscillators and their bath temperature constrain an entanglement
generation protocol and its outcome. In this paper, we show that the above analytic treatment can be achieved by choosing a {\em specific form} of the qubit-oscillator interaction which is naturally available in quantum nanomechanics and can be implemented with minimal effort in cavity/circuit-QED. Together with the inclusion of a thermal environment for the oscillator, our treatment also enables the analytic inclusion of the dephasing of qubits that may be important in a quantum nanomechanical setting. The scheme we are considering for the entanglement preparation, relying on the strength of
resonant interactions, can be faster than that with cross-Kerr-like interactions \cite{Ralph}. We also find an allowed time window for the
entanglement to appear which becomes broad when decoherence and temperature are low. This can make the scheme more robust to errors in
interaction times than most other schemes \cite{Davidovich,Browne}. In addition, we show the possibility of reciprocation of the entanglement of
the oscillators to two qubits. Thus the generated mixed state of the oscillator can serve as a resource for qubit based quantum information processing (QIP).

Our results should be particularly useful for quantum nanomechanics, where oscillators
are yet to be cooled to
  the ground state.
In technologies where preparation of a ground state oscillator is available \cite{Haroche1,Blais}, using thermal states will surely lower the
demands on cooling and isolation. Our scheme will show that oscillators in almost classical states can indeed be entangled to provide a QIP
resource even when they are continuously (but weakly) coupled to a thermal environment.
 
 In addition to this, we believe the solution method presented here to be relevant on its own, as it could be applied to a variety of problems involving the same interaction Hamiltonian and the same decoherence model.
\section*{PROBLEM DESCRIPTION}

   We start by presenting the schematics of our entanglement generation protocol without specifying the Hamiltonian. At first, we
let two identical oscillators interact with the same qubit, which we call the \textit{entangling qubit}, either sequentially (such as for an atomic qubit flying
through a pair of cavities) or simultaneously (such as for a superconducting qubit). In the former case we will assume the two interaction times to be the same. The
states of the qubit are $\{\ket g,\ket e\}$, while the oscillators are described by the annihilation operators $a,b$. Let the oscillators be initially in a thermal state and
the qubit in $\ket g$ so that the initial state of the complete system is:
\begin{align}
&\rho(0)=\kebra{g}{g}\otimes\rho_\textrm{th}=\kebra{g}{g}\otimes\frac{e^{-\beta a^\dagger a}e^{-\beta b^\dagger b}}{Z^2},
\end{align}
where $Z=\textrm{Tr}\{e^{-\beta a^\dagger a}\}=\textrm{Tr}\{e^{-\beta b^\dagger b}\}$ and $\beta=\omega/(k_B T)$. Throughout the paper, the unit of $\hbar=1$ is assumed. After an interaction between
the qubit and the oscillators for a time $t$ the system will evolve to $\rho(t)$. At this point, we measure the internal state of the qubit to
obtain two possible outcomes $f=g,e$. As a result the oscillators are projected in the mixed state $\rho_f(t)$. Our aim is to show that for an
appropriate interaction, such a state may be entangled by an amount which depends on the temperature $T$ and the strength of decoherence.
 One could try to detect the entanglement contained in the state $\rho_f$, for example by verifying a Bell inequality violation
 by local measurements on the oscillators (we will discuss how to do this for our specific model).
Alternatively, one could extract part of the entanglement on to an additional pair of qubits $A$ and $B$ (entanglement ``reciprocation"),
initially in the separable state $\ket{g,g}=\ket g_A\otimes\ket g_B$ by making mode $a$ interact with qubit $A$ and mode $b$ with qubit $B$
respectively. The resulting operation alone may not suffice to entangle $A$ and $B$, as they will, in general, remain entangled with modes $a$
and $b$. So we perform additional measurements of momenta $\hat p_a=-i(a-a^\dagger);\quad \hat p_b=-i(b-b^\dagger)$ or the parities
$\hat\Pi_a=(-1)^{a^\dagger a};\quad \hat\Pi_b=(-1)^{b^\dagger b}$ of the two oscillators. Note that no entanglement is created during such reciprocation
since the two parts of the system ($A,a$ and $B,b$) are not directly interacting and the measurement is performed locally. Thus
measuring $A$ and $B$ to be entangled is a \textit{sufficient condition} for the state $\rho_f(t)$ to be entangled. It also demonstrates that
the oscillators can store entanglement to be later extracted by qubits.
\section*{SOLUTION METHODS}
  We present the form of the qubit-oscillator interaction we use, first showing the case of a single oscillator and then generalizing to the
case of two oscillators.
The qubit and the oscillator are resonant at frequency $\omega$. The Hamiltonian in the interaction picture is:
\begin{equation}
H=\sigma_1(a+a^\dagger),\label{hamiltonian}
\end{equation}
where we have introduced the Pauli operators
$\sigma_1=\kebra{e}{g}+\kebra{g}{e};\sigma_2=i(\kebra{e}{g}-\kebra{g}{e});\sigma_3=\kebra{g}{g}-\kebra{e}{e}$. This Hamiltonian can arise in a
Jaynes-Cummings system by driving it with an external laser \cite{Agarwal-Solano} (this {\em revives the counter-rotating terms} while
preserving the strength of the coupling \cite{Agarwal-Solano}). It also arises naturally in nanomechanical oscillators coupled to charge qubits (although
in the latter case, a rotation of the qubit basis such that $\sigma_3\rightarrow\sigma_1$ is needed). The time evolution operator is
\begin{align}
&U(t)=e^{-iHt}=D(-i\sigma_1t)
\end{align}
where $D(\alpha)=e^{\alpha a^\dagger-\alpha^*a}$ is the displacement operator \cite{walls-milburn} (we can use $\sigma_1$ inside this operator as
$[\sigma_1a,\sigma_1a^\dagger]=[a,a^\dagger]=1$). If the qubit is interacting with two oscillator modes $a,b$ the previous expression is modified into:
\begin{equation}
U(t)=D_a(-i\sigma_1t)D_b(-i\sigma_1t),
\end{equation}
where $D_i$ is the displacement operator for mode $i$ \cite{walls-milburn}. If the entangling qubit is measured in $\ket g$ or $\ket e$, the resulting operators
acting on the oscillators (combining evolution and measurement) are
\begin{align}
&\bra g U(t)\ket g=\tfrac{1}{2}(D_a(it)D_b(it)+D_a(-it)D_b(-it)),\nonumber\\
&\bra e U(t)\ket g=-\tfrac{1}{2}(D_a(it)D_b(it)-D_a(-it)D_b(-it)).\label{effective}
\end{align}
The above operators have increasing entangling capabilities as $t$ gets larger, as can be seen for example by applying one of them to a product of two coherent states.\\
At this point we include the effect of a thermal environment in our treatment, for which we first get back to the interaction of a single oscillator with a single qubit. Introducing decoherence, the dynamics can be described by the master equation:
\begin{equation}
\frac{\partial \rho}{\partial t}=-i[\sigma_1(a+a^\dagger),\rho]+L(T)\rho+L_\phi\rho,
\end{equation}
where $L(T)$ is the Lindblad operator for the oscillator in a thermal bath at temperature $T$, which will, for cases we consider, be of the amplitude
damping form \cite{walls-milburn}:
\begin{align}
&L(T)\rho=\frac{\kappa}{2}(n(T)+1)(2a\rho a^\dagger-a^\dagger a\rho-\rho a^\dagger a)+\nonumber\\
&\phantom{L(T)\rho=}+\frac{\kappa}{2}n(T)(2a^\dagger\rho a-a a^\dagger\rho-\rho a a^\dagger),
\end{align}
where $n(T)=(exp(\beta)-1)^{-1}$ and $\kappa$ is the damping rate. The operator $L_\phi\rho=\frac{\gamma}{2}(\sigma_1\rho\sigma_1-\rho)$ will be present for the case of charge qubits whose dephasing can be significant (here $\sigma_1$ instead of $\sigma_3$ has been used because for a charge qubit the basis is rotated, as mentioned earlier). The model is analytically solvable, and we think it is worth to present here the solution method in full detail, as it may be applied to other problems involving the same form of interaction Hamiltonian. We switch to the Wigner representation of the oscillator defined as
\begin{equation}
W(\alpha,t)=\frac{1}{\pi^{2}}\int d^2\eta\phantom{0} e^{\eta^*\alpha-\eta\alpha^*}\textrm{Tr}\{D(\eta)\rho(t)\}.
\end{equation}
This representation is widely used in quantum optics, where the density matrix of a single (or multi) mode field is encoded into the real-valued Wigner function \cite{walls-milburn}.
However, since in this paper we are dealing with a composite system (oscillator + qubit), the state of the system can not be represented by a single function; it is indeed encoded in a {\em $2\times2$ matrix of functions}, which we
may call the {\em Wigner matrix}. The Wigner matrix component $W_{i,j}$ is defined as the Wigner function for the operator $\langle i|\rho(t)\ket j$ where $i,j=e,g$. Note that $W_{i,j}$ is not necessarily real valued for $i\neq j$; nevertheless if $\rho$ is a genuine density matrix then the condition $W_{i,j}=W_{j,i}^*$ holds, i.e. the Wigner matrix is hermitian for any $(\alpha,t)$.
The master equation for $\rho$ can be converted into a differential equation for $W$ by using the following correspondences (for a rigorous derivation see for example \cite{walls-milburn}):
\begin{align}
&a\rho\rightarrow \left(\alpha+\frac{1}{2}\partial_{\alpha^*}\right)W;\qquad\rho a\rightarrow \left(\alpha-\frac{1}{2}\partial_{\alpha^*}\right)W;\\
&a^\dagger\rho\rightarrow \left(\alpha^*-\frac{1}{2}\partial_{\alpha}\right)W;\qquad\rho a^\dagger\rightarrow \left(\alpha^*+\frac{1}{2}\partial_{\alpha}\right)W.
\end{align}
It follows that the Wigner matrix obeys the differential equation
\begin{align}
&\partial_t W=-i(\alpha+\alpha^*)[\sigma_1,W]-\frac{i}{2}(\partial_{\alpha^*}-\partial_\alpha)\{\sigma_1,W\}+\nonumber\\
&\phantom{\partial_t W}+L(T)W+L_\phi W,\label{diffeq}
\end{align}
where we introduced the anticommutator $\{o_1,o_2\}=o_1o_2+o_2o_1$.
The Lindblad operator in terms of the complex variable $\alpha$ is
\begin{equation}
L(T)W=\frac{\kappa}{2}\left(\partial_\alpha \alpha+\partial_{\alpha^*}\alpha^*+2\Delta(T)\partial_\alpha\partial_{\alpha^*}\right)W,
\end{equation}
where $\Delta(T)=n(T)+1/2$. Note that (\ref{diffeq}) is indeed a system of coupled differential equations for the four components of the matrix $W$, which can be separated by choosing an appropriate decomposition of $W$ in the Pauli basis $\{1,\sigma_1,\sigma_2,\sigma_3\}$. In fact we can see that the commutator $[\sigma_1,\cdot]$ in (\ref{diffeq}) is non-vanishing only if it is applied to the Pauli operators $\sigma_2,\sigma_3$, while the anticommutator $\{\sigma_1,\cdot\}$ is non-vanishing only when it is applied to the operators $1,\sigma_1$. It is then a simple step further to verify that
\begin{align}
&[\sigma_1,\sigma_2\pm i\sigma_3]=\pm(\sigma_2\pm i\sigma_3),\\
&\{\sigma_1,1\pm\sigma_1\}=\pm(1\pm\sigma_1).
\end{align}
We have at this point all the tools we need to write down a decomposition of the Wigner matrix in ''normal modes``
\begin{align}
&W=\tfrac{1}{4}(v_1(1+\sigma_1)+v_2(1-\sigma_1)+\nonumber\\
&\phantom{W}+v_3(\sigma_2-i\sigma_3)+v_4(\sigma_2+i\sigma_3)),
\end{align} 
which then yields four decoupled differential equations:
\begin{align}
&\partial_t v_1=-\frac{i}{2}(\partial_{\alpha^*}-\partial_\alpha)v_1+L(T)v_1,\label{v1}\\
&\partial_t v_2=\frac{i}{2}(\partial_{\alpha^*}-\partial_\alpha)v_2+L(T)v_2,\\
&\partial_t v_3=2i(\alpha+\alpha^*)v_3+L(T)v_3-\gamma v_3,\label{w1}\\
&\partial_t v_4=-2i(\alpha+\alpha^*)v_4+L(T)v_4-\gamma v_4.\label{diffeqs}
\end{align}
We suppose that initially the oscillator is at thermal equilibrium with its environment at temperature $T$, while the qubit is in the ground state. In terms of initial conditions this translates into
\begin{align}
&v_1(0)=v_2(0)=-iv_3(0)=iv_4(0)=W_T(\alpha),\label{initials}\\
&W_T(\alpha)=\frac{1}{\pi \Delta}e^{-\frac{|\alpha|^2}{\Delta}}.
\end{align}
These initial conditions allow for a compact solution which we are going to present in full, while the solution for generic initial conditions can be expressed in terms of convolution integrals as discussed in Appendix.
Equation (\ref{v1}) can be solved by choosing an ansatz of the form
\begin{align}
&v_1(\alpha,t)=W_T(\alpha+\lambda(t)).\label{ansv1}
\end{align}
We know that $W_T(\alpha)$ is such that $L(T)W_T(\alpha)=0$, and clearly $\partial_tW_T(\alpha)=0$. Thus, if (\ref{ansv1}) holds, it follows that
\begin{align}
&L(T)v_1=-\frac{\kappa}{2}\left(\lambda\partial_{\alpha}+\lambda^*\partial_{\alpha^*}\right)v_1,\\
&\partial_tv_1=\left(\dot\lambda\partial_\alpha +\dot\lambda^*\partial_{\alpha^*}\right)v_1.\label{dtv1}
\end{align}
By substituting these into Eq. (\ref{v1}) and collecting terms proportional to $\partial_\alpha v_1$, we obtain the differential equation
\begin{align}
&\dot\lambda=\frac{i}{2}-\frac{\kappa}{2}\lambda,\label{lambda}
\end{align}
with initial condition $\lambda(0)=0$ as a consequence of (\ref{initials}). Such equation can be easily solved, yielding
\begin{equation}
\lambda(t)=\frac{i}{\kappa}\left(1-e^{-\frac{\kappa}{2}t}\right).
\end{equation}
Following the same procedure we obtain
\begin{equation}
v_2(\alpha,t)=W_T(\alpha-\lambda(t)).
\end{equation}
For $v_3$ we consider the following ansatz:
\begin{equation}
v_3(\alpha,t)=iW_T(\alpha)e^{\mu(t)(\alpha+\alpha^*)+\nu(t)}.
\end{equation}
Similar considerations as before allow us to express
\begin{align}
&L(T)v_3=\frac{\kappa}{2}\left(-(\alpha+\alpha^*)\mu+2\Delta \mu^2\right)v_3,\\
&\partial_t v_3=\left((\alpha+\alpha^*)\dot\mu+\dot\nu\right)v_3.
\end{align}
This time, after substituting such expressions in Eq. (\ref{w1}), we can collect separately the terms proportional to $\alpha v_3$ and those proportional to $v_3$, to obtain the pair of coupled differential equations
\begin{align}
&\dot \mu=2i-\frac{\kappa}{2}\mu,\label{mu}\\
&\dot \nu=\kappa\Delta \mu^2-\gamma.\label{nu}
\end{align}
From (\ref{initials}) we get the initial conditions $\mu(0)=\nu(0)=0$, then the solutions of Eqs. (\ref{mu}) and (\ref{nu}) are
\begin{align}
&\mu(t)=4\lambda(t),\\
&\nu(t)=-\gamma t+\kappa\Delta\int_0^t\mu(\tau)^2d\tau.
\end{align}
Again, $v_4$ can be solved similarly, yielding
\begin{equation}
v_4(\alpha,t)=-iW_T(\alpha)e^{-\mu(t)(\alpha+\alpha^*)+\nu(t)}.
\end{equation}

 At this point we can go back to our original problem and generalize the procedure for a qubit interacting with two oscillators. We have in this case a master equation for the two-mode Wigner matrix $W(\alpha,\beta)$.
If we suppose that the two oscillators have the same temperature and the same value of $\kappa$, the solution is obtained from the single mode case by replicating each function, i.e. $v_i(\alpha,t)\rightarrow v_i(\alpha,t)v_i(\beta,t)$. Note that this last operation implies that the qubit is interacting simultaneously with the two oscillators. If the interaction is sequential we should consider the delay between the two interactions, which makes the first oscillator decay for longer (dephasing can usually be neglected for flying qubits). It is not hard to include this effect analytically if needed, however our calculations show that it can be taken into account with good approximation by doubling the damping rate of the oscillators, so that we can present a common treatment for the two settings. The Wigner function of the two oscillators after the qubit is measured in $f=e,g$ is
\begin{align}
&W_f(\alpha,\beta,t,T)=(\mathcal N(t,T))^{-1}\bra f W(\alpha,\beta,t,T)\ket f,\label{oscillstate}
\end{align}
where $\mathcal N(t,T)=\int d^2\alpha d^2\beta\phantom{o}\bra fW(\alpha,\beta,t,T)\ket f$.
  
  In the reciprocation procedure the two qubits plus
two oscillators are described by a 4x4 Wigner matrix, initially given by $\kebra{gg}{gg}\times W_f(\alpha,\beta,t,T)$, and then evolved for
a further time $t$ (for simplicity the same as before). Similar techniques as above allow to keep the treatment analytical. The state of the qubits after reciprocation is obtained from the Wigner matrix in both cases of our interest. For momentum measurements we use the property
\begin{equation}
\bra{p_a,p_b}\rho\ket{p_a,p_b}=\int dx_a dx_b W(x_a+ip_a,x_b+ip_b),
\end{equation}
which gives the overlap of the density matrix with the momentum eigenstates. In this case we decompose the complex variables into their respective real and imaginary parts, i.e. $\alpha=x_a+ip_a$, $\beta=x_b+ip_b$. When measuring parities we use
\begin{equation}
W(\alpha,\beta)=\frac{4}{\pi^2}\textrm{Tr}\{D_a(\alpha)^\dagger D_b(\beta)^\dagger\rho D_a(\alpha)D_b(\beta)\hat\Pi_a\hat\Pi_b\},
\end{equation}
 which is a well known alternative definition \cite{walls-milburn}. For example, the two qubit density matrix corresponding to the parity eigenvalues $\Pi_a=\Pi_b=+1$ is
\begin{align}
&\rho_{++}(t)\propto\int d^2\alpha d^2\beta W(\alpha,\beta)+\frac{\pi}{2}\int d^2\alpha W(\alpha,0)+\nonumber\\
&\phantom{\rho_{++}(t)\propto}+\frac{\pi}{2}\int d^2\beta W(0,\beta)+\frac{\pi^2}{4}W(0,0).
\end{align}
\section*{ENTANGLEMENT ANALYSIS}

  We proceed to evaluate the entanglement contained in the states prepared at each stage of the protocol. Let us start with state $\rho_f$ of
the two oscillators after the entangling qubit is measured. We investigate its nonlocal properties by looking for a Bell inequality
violation. We consider the measurable quantity
\begin{equation}
\Delta P(\alpha,\beta)=\frac{\pi^2}{4}W_f(\alpha,\beta)= P_{\textrm{same}}(\alpha,\beta)-P_{\textrm{diff}}(\alpha,\beta),
\end{equation}
where $P_\textrm{same/diff}(\alpha,\beta)$ is the probability of measuring the two oscillators with same/opposite parities after having applied
the displacements $D_a(\alpha)^\dagger, D_b(\beta)^\dagger$. This quantity takes values between -1 and 1 and therefore it can be used to form a Bell inequality:
\begin{align}
&\mathcal B(\alpha,\beta,\alpha',\beta')=\Delta P(\alpha,\beta)+\Delta P(\alpha',\beta)+\nonumber\\
&.\phantom{\mathcal B(\alpha,\beta,\alpha',\beta')=}+\Delta P(\alpha,\beta')-\Delta P(\alpha',\beta')\leq2,
\end{align}
where $\alpha,\beta,\alpha',\beta'$ are arbitrary complex numbers. A value of $\mathcal B$ greater than 2 indicates the presence of quantum nonlocal correlations. The maximum violation for a bipartite system is $\mathcal B=2\sqrt2$, corresponding to strong nonlocality. In the limit of large $t$ and $\kappa,\gamma\rightarrow0$, the maximum value of $\mathcal B$ is given by
\begin{equation}
\mathcal B_{\textrm{max}}\simeq\frac{1}{(1+2n(T))^2}2\sqrt2,
\end{equation}
corresponding to a set of solutions $\{\bar \alpha,\bar \beta,\bar \alpha',\bar \beta'\}$.
\begin{figure}[t!]
\includegraphics[width=0.24\textwidth]{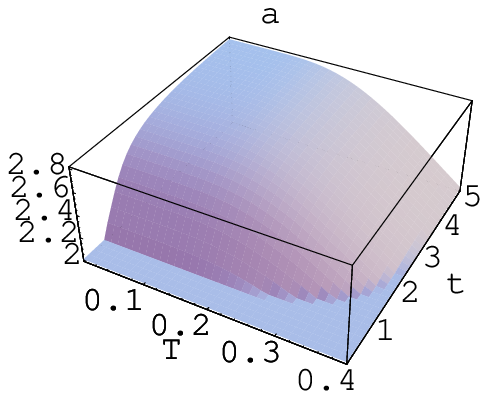}\includegraphics[width=0.24\textwidth]{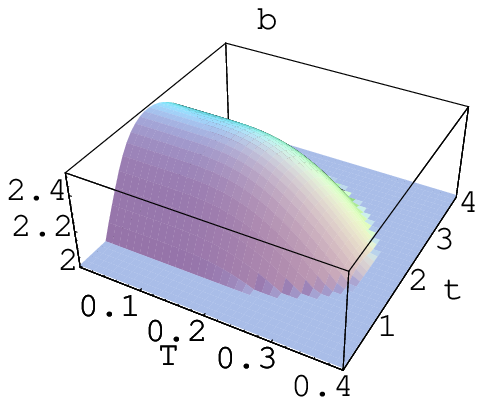}
\caption{Lower bound for the Bell inequality violation $\textrm{Max}(\mathcal B,2)$, relative to state $\rho_g$ as a function of temperature and interaction time. The case $\rho_e$ is qualitatively similar. Plot (a) corresponds to the ideal case $\gamma,\kappa\sim0$, where the maximum violation compatible with the system's temperature is reached for long enough times. In plot (b) $(\kappa,\gamma)=(5,3)\times10^{-3}$ as in \cite{Blais}, which limits the time window to circa 2 Rabi periods and makes the maximum violation drop to approximately 2.4\label{firstfig}}
\end{figure}
We can see that if $T$ is lower than a critical temperature $T_c\simeq0.408\omega$, then a violation of the Bell inequality holds. In the case of finite times and $\kappa,\gamma>0$, we get a lower bound for the Bell inequality violation by considering
\begin{equation}
\mathcal B_\textrm{max}\geq\mathcal B(\bar \alpha,\bar \beta,\bar \alpha',\bar \beta'),
\end{equation}
where $t$ has to be substituted by $\frac{2}{\kappa}(1-e^{-\kappa t/2})$. By introducing decoherence the time window in which it is possible to verify a Bell inequality violation becomes limited and the maximum value of $\mathcal B$ drops from the ideal value of $2\sqrt2$. Nevertheless
if the decoherence and dephasing rates are reasonably small we are still left with a range of times and temperatures which give a value of $\mathcal B$
significantly larger than 2, as we can see in Fig.~\ref{firstfig}.

\begin{figure}[H]
\includegraphics[width=0.24\textwidth]{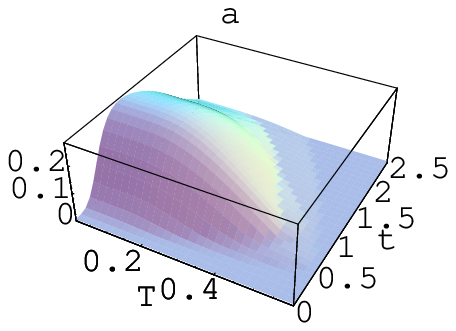}\includegraphics[width=0.24\textwidth]{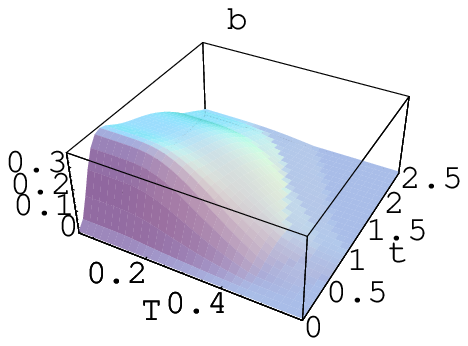}
\caption{Average negativity of the two-qubit density matrix in the case of parity measurements.  Plot (a) refers to the entangling qubit being measured in $g$, while plot (b) refers to the outcome $e$. The maxima are close to the value $\sim0.3$. $(\kappa,\gamma)=(0.007,0)$ as in \cite{Haroche1}
\label{thirdfig}}
\end{figure}

 Let us discuss our results for the reciprocation procedure.
\begin{figure}[t!]
\includegraphics[width=0.16\textwidth]{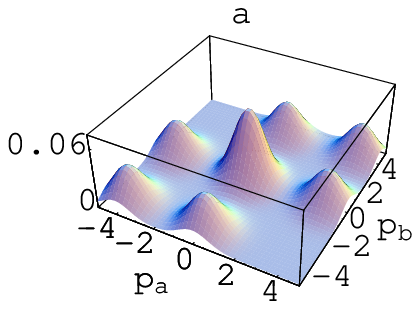}\includegraphics[width=0.16\textwidth]{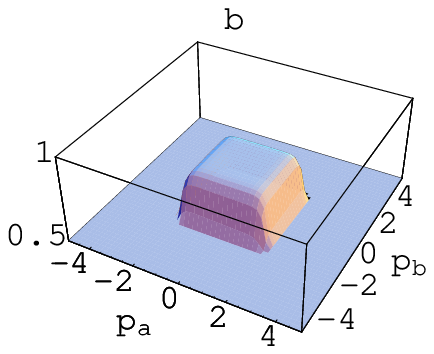}\includegraphics[width=0.16\textwidth]{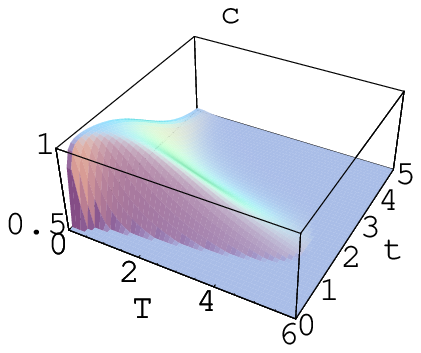}
\caption{Entanglement reciprocation via momentum measurements, for the entangling qubit measured in $e$. The $g$ case is qualitatively similar.
In plots (a) and (b) time and temperature have been fixed ($t=2$ and $T=\omega$). Plot (a) shows the probability
distribution for the outcomes $p_a,p_b$. Plot (b) shows the fidelity of the two qubit density matrix to the Bell state $\ket{\psi^+}=\tfrac{1}{\sqrt2}(\ket{ge}+\ket{eg})$. A fidelity above 0.5 indicates an entangled state.
Entanglement is conditioned to measuring $p_a,p_b$ inside the central peak of the probability distribution, which has a volume $\sim$25\%. The remaining peaks yield separable qubit states. In plot (c) the maximum fidelity (corresponding to $p_a,p_b\simeq0$) is shown as a function of temperature and interaction time. We set $\kappa=\gamma=0.01$ in all plots.
\label{secondfig}}
\end{figure}
In the case of momentum measurements we can find, conditionally, the qubits in an entangled state for values of $T$ well above $T_c$, as shown
in Fig.~\ref{secondfig}. Remarkably this procedure sustains higher and higher temperatures as the parameters $\kappa$ and $\gamma$ approach
zero. For $\kappa,\gamma=0$ we found no temperature limit at all, and we only had to choose long enough interaction times to let the
entanglement appear. Fig.\ref{thirdfig} shows the entanglement (negativity) of the two-qubit density matrix in the case of reciprocation with
parity measurements, averaged over the four possible outcomes $\Pi_a,\Pi_b=\pm1$. In this case we found again a limit in the sustainable
temperature, which does not increase as $\kappa,\gamma\rightarrow0$. The presence of a maximum allowed temperature may be a peculiarity of
parity measurements. It is important to point out that the interaction time does not need to be finely tuned at any step of our protocol,
provided that we remain in the proper time window which allows the entanglement to appear. In all the situations presented this time window
increases indefinitely as we increase our ability to build systems with smaller $\kappa$ and $\gamma$, so that technological advances in
building higher quality qubits and oscillators will eventually remove any need of fine tuning in the interaction times.
\section*{POSSIBLE IMPLEMENTATIONS AND CONCLUSIONS}
 Presently cavity-QED setups with flying atoms seem to be a promising ground for the implementation of our scheme, e.g. in \cite{Haroche1}
Rauschenbeutel {\em et al.} achieved $(\kappa,\gamma)\simeq(0.007,0)$. Circuit-QED setups with superconducting qubits are also very appealing since values of
$(\kappa,\gamma)\simeq(5,3)\times10^{-3}$ or even better can be achieved by present technology \cite{Blais}.
Values of $(\kappa,\gamma)\simeq(0.23,0.005)$, which do not allow for a Bell inequality violation but still give decent
results for reciprocation (e.g a maximum Bell-state fidelity above 0.85), seem feasible at present for a spin coupled to a nanomechanical resonator,
 as shown in \cite{Treutlein}. In the context of quantum nanomechanical systems coupled to superconducting qubits $\kappa\sim0.01$ is within reach, but the rapid
 qubit dephasing $\gamma\sim1$ is still an issue for present technology, although some progress seems possible in the near future \cite{LaHaye}.
 Note that parity measurements have already been performed successfully in a cavity-QED setting \cite{measure1}, while momentum measurements should be
 possible with capacitive transducers for quantum nanomechanics.

 We are convinced that the robustness of the protocol relies on the features of the Hamiltonian (\ref{hamiltonian}). In fact, as we see in (\ref{effective}), it allows us to apply to the oscillators a coherent superposition of displacements in opposite directions, with amplitude proportional to $t$. We expect to see entanglement between the two oscillators if $t$ is larger than the phase space extension of the initial state, even if mixed. However, large $t$ also means more decoherence, thus the competition between these two effects will determine the existence and extension of a time window in which entanglement can be established.
 
  We thank the Engineering and Physical Sciences Research Council United Kingdom, the
Quantum Information Processing Interdisciplinary Research Collaboration, the Royal Society and the Wolfson Foundation.
\appendix*
\section{Solution for generic initial conditions via the integral kernels} 
The integral Kernels $V_j(\alpha,\alpha_0,t);$ $j=1,..,4;$ are solutions of the equations (\ref{v1}-\ref{diffeqs}) with initial condition
\begin{equation}
V_j(\alpha,\alpha_0,0)=\delta(\alpha-\alpha_0).
\end{equation}
From the integral kernel the solution corresponding to an arbitrary initial condition $v_j(\alpha,0)$ can be expressed in terms of a convolution integral:
\begin{equation}
v_j(\alpha,t)=\int  V_j(\alpha,\alpha_0,t)v_j(\alpha_0,0)d\alpha_0
\end{equation}
It is possible to derive the four integral Kernels as follows. We know from the literature (see e.g. \cite{walls-milburn}) that the function
\begin{equation}
K_0(\alpha,\alpha_0,t)=\frac{1}{\pi \Delta \left(1-e^{-\kappa t}\right)}e^{-\frac{\left|\alpha-\alpha_0 e^{-\kappa t/2}\right|^2}{\Delta \left(1-e^{-\kappa t}\right)}}
\end{equation}
is the integral kernel for $L(T)$, i.e.
\begin{align}
&\partial_tK_0(\alpha,\alpha_0,t)=L(T)K_0(\alpha,\alpha_0,t),\\
&K_0(\alpha,\alpha_0,0)=\delta(\alpha-\alpha_0)
\end{align}
To find the four integral kernels of equations (\ref{v1}-\ref{diffeqs}), we consider the four ansatz
\begin{align}
&V_1(\alpha,\alpha_0,t)=K_0(\alpha+\lambda(t),\alpha_0,t)\\
&V_2(\alpha,\alpha_0,t)=K_0(\alpha-\lambda(t),\alpha_0,t)\\
&V_3(\alpha,\alpha_0,t)=K_0(\alpha,\alpha_0,t)e^{\mu(t)(\alpha+\alpha_0+\alpha^*+\alpha_0^*)+\nu(t)}\label{ker3}\\
&V_4(\alpha,\alpha_0,t)=K_0(\alpha,\alpha_0,t)e^{-\mu(t)(\alpha+\alpha_0+\alpha^*+\alpha_0^*)+\nu(t)}\label{ker4}
\end{align}
Using the techniques previously described, we see that the first two ansatz lead again to equation (\ref{lambda}), which we already solved. For the last two ansatz to be actual solutions, the following differential equations must be satisfied:
\begin{align}
&\dot\mu=2i-\frac{\kappa}{2}\frac{1+e^{-\kappa t}}{1-e^{-\kappa t}}\mu\\
&\dot\mu=\kappa\frac{e^{-\kappa t/2}}{1-e^{-\kappa t}}\mu\\
&\dot\nu=\kappa\Delta \mu^2-\gamma.
\end{align}
These are obtained by substituting (\ref{ker3}) and (\ref{ker4}) into (\ref{w1}) and (\ref{diffeqs}) respectively, then collecting separately the terms proportional to $V_j,\alpha V_j$ and $\alpha_0V_j$, with $j=3,4$.
The system of equations above is overdetermined in general, however it is possible to find a particular solution which also accounts for the initial conditions $\mu(0)=\nu(0)=0$:
\begin{align}
&\mu(t)=\frac{4i}{\kappa}\frac{1-e^{-\kappa t/2}}{1+e^{-\kappa t/2}},\\
&\nu(t)=\kappa\Delta\int_0^t\mu(\tau)^2d\tau-\gamma t.
\end{align}
We have at this point all the tools needed to describe the evolution of any given initial state in terms of integrals. The procedure is easily generalizable (at least in principle) to an ensemble $\mathcal Q=\{q_1,...,q_k\}$, containing $k$ qubits, and an ensemble $\mathcal A=\{a_1,...,a_l\}$ containing $l$ oscillators, such that each qubit $q_j$ is coupled to a subset of $\mathcal A$ with the Hamiltonian (\ref{hamiltonian}), while objects of the same kind do not interact.

\end{document}